\documentclass[aps,pra,twocolumn,nopacs,superscriptaddress,floatfix]{revtex4-1} 

\usepackage{amsmath} \usepackage{bbm}
\usepackage{epsfig}
\usepackage{color}
\usepackage{graphicx}
\usepackage{epstopdf}

\newcommand{\id}{\mathbbm{1}}

\usepackage{times}

\newcommand{\cc}{{\mathbbm{C}}}
\newcommand{\rr}{{\mathbbm{R}}}
\newcommand{\nn}{{\mathbbm{N}}}

\newtheorem{theorem}{Theorem}

\newtheorem{definition}[theorem]{Definition}

\begin{document}

\title{Directly estimating non-classicality}

\author{A.\ Mari}
\author{K.\ Kieling} 

\affiliation{Institute of Physics and Astronomy, University of Potsdam, 14476 Potsdam, Germany}

\author{B.\ Melholt Nielsen}
\author{E.\ S.\ Polzik}

\affiliation{Niels Bohr Institute, Danish National Research Foundation Center
for Quantum Optics, DK-2100 Copenhagen, Denmark}

\author{J.\ Eisert}

\affiliation{Institute of Physics and Astronomy, University of Potsdam, 14476 Potsdam, Germany}
\affiliation{Institute for Advanced Study Berlin, 14193 Berlin, Germany}

\begin{abstract}
We establish a method of directly measuring and estimating
non-classicality---operationally defined in terms of the 
distinguishability of a given state from one with a positive Wigner function.
It allows to certify non-classicality, based on possibly much fewer measurement settings than 
necessary for obtaining complete tomographic knowledge, 
and is at the same time equipped with a full certificate. We find 
that even from measuring two conjugate variables alone, one may infer 
the non-classicality of quantum mechanical modes. This method also
provides a practical tool to eventually certify such features in mechanical degrees
of freedom in opto-mechanics. The proof of the result is based on 
Bochner's theorem characterizing classical and quantum characteristic functions
and on semi-definite programming. In this joint theoretical-experimental work we 
present data from experimental optical Fock state preparation, demonstrating 
the functioning of the approach. 
\end{abstract}	

\maketitle

Where is the ``boundary'' between classical and quantum physics? Unsurprisingly,
acknowledging that 
quantum mechanics is the fundamental theory from which classical properties
should
emerge in one way or the other, instances of this question have a long tradition
in physics. 
Possibly the most conservative and stringent criterion for 
non-classicality of a quantum 
state of bosonic modes is that the Wigner function---the closest analogue to a
classical probability 
distribution in phase space---is negative, and can hence no longer be
interpreted as a classical 
probability distribution \cite{Wigner,Leonhardt,Zyc}.
From this, negativity of other quasi-probability distributions, familiar in quantum optics, 
such as the $P$-function \cite{Wigner,Vogel} follows. 
In fact, a lot of experimental progress was
made in recent years on 
preparing quantum states of light modes that exhibit such non-classical
features, when preparing number 
states, photon subtracted states, or small Schr{\"o}dinger cat states
\cite{Rauschenbeutel,Ourjoumtsev06,Polzik}. At the same time, a lot of effort is
being made of driving mesoscopic mechanical degrees of freedom into quantum
states 
eventually showing such non-classical features \cite{Optomechanics}. 
All this poses the question, needless to say, of how to best and 
most accurately certify and measure such features.

In this work, (I) we demonstrate that, quite remarkably, non-classicality in the
above sense can be detected from
mere measurements of two conjugate variables. For a single mode, this amounts to
position and momentum
detection, as can be routinely done by homodyne measurements in optical systems.
(II) What is more, using such data (or also data that are tomographically complete)
one can get a direct and rigorous lower  bound to the probability of operationally distinguishing 
this quantum state from one with a positive Wigner function---including a full certificate.
Such a bound uses information from possibly much
fewer measurement settings than needed for full quantum state
tomography. At the same time, quantum state tomography
using Radon transforms for quantum modes is overburdened with  problems of
ill-conditioning: This gives, 
strictly speaking, rise to the situation that when fully reconstructing a state
based on such tomographically
complete data, one often should expect to encounter such large error bars that
the resulting state would well be also 
consistent with having had no non-classicality at all. 

The method introduced here, in contrast, is a {\it direct method} giving rise to a {\it certified 
bound} which arises from 
conditions all classical and quantum characteristic functions have to satisfy as
being grasped by the classical and quantum Bochner's theorem \cite{Hudson}. 
Hence, we ask: {\it ``What is the least non-classical state
consistent with the data''}? Intuitively speaking, the proof circles around the
deviation of a quantum
characteristic function as the Fourier transform of the Wigner function from a
classical characteristic function.
This deviation can then be formulated in terms of a semi-definite program---so
a well-behaved convex
optimization problem---giving rise to certifiable bounds. The same
technique can also be applied
to notions of entanglement, and indeed, the rigor applied here reminds of 
applying quantitative entanglement witnesses. 
In this sense, one can directly certify quantum
properties with much less
than full tomographic knowledge. What is more, the criterion evaluation procedure is efficient.  
At present, such techniques should be most applicable to 
systems in quantum optics, and we indeed implement this idea in a {\it quantum 
optical experiment} preparing a field mode in a non-classical state. 
Yet, they should be expected to be helpful when eventually
certifying that a mesoscopic mechanical system has eventually reached quantum properties
\cite{Optomechanics},
where ``having achieved a non-classical state'', with careful error analysis, 
will constitute an important benchmark.

{\it Measure of non-classicality. --} 
Non-classicality is most reasonably quantified in terms of the 
possibility of operationally distinguishing a given state from a
state that one would conceive as being classical.  That is to say,
the meaningful notion of distinguishing a state from a classical one is as
follows.

\begin{definition}[Measure of non-classicality] Non-classicality is measured
in terms of the operational distinguishability of a given state from a 
state having a positive Wigner function,
\begin{equation}
\eta( \rho)=\min_{ \omega \in \mathcal C}\| \rho- \omega \|_{1}
\end{equation}
where $\mathcal C$ denotes the set of all quantum states with positive Wigner 
function and $\|\cdot \|_1$ is the trace norm. 
\end{definition}

This measure is indeed {\it the} operational definition of
a non-classical state---as long as one accepts the negativity of the Wigner
function as the figure of merit of non-classicality. Needless to say, the 
operational distinguishability with respect to other properties would also
be quantified by trace-distances, and naturally several quantities
of such a type can be found in the literature (see, e.g., Ref.\ \cite{measure}).
By definition, it has among others the following natural
properties:

(a) It is invariant under passive and active linear transformations.

(b) It is non-increasing under Gaussian channels, and in fact under any
operation
that cannot map a state with a positive Wigner function onto a negative one.

The latter property is an immediate consequence of the trace norm being
contractive under completely positive maps. 
Moreover since Gaussian states are positive this measure of negativity gives
also a direct lower bound to the non-Gaussianity of the same state.

\begin{figure}[tbh]
\centerline{\includegraphics[width=0.42 \textwidth]{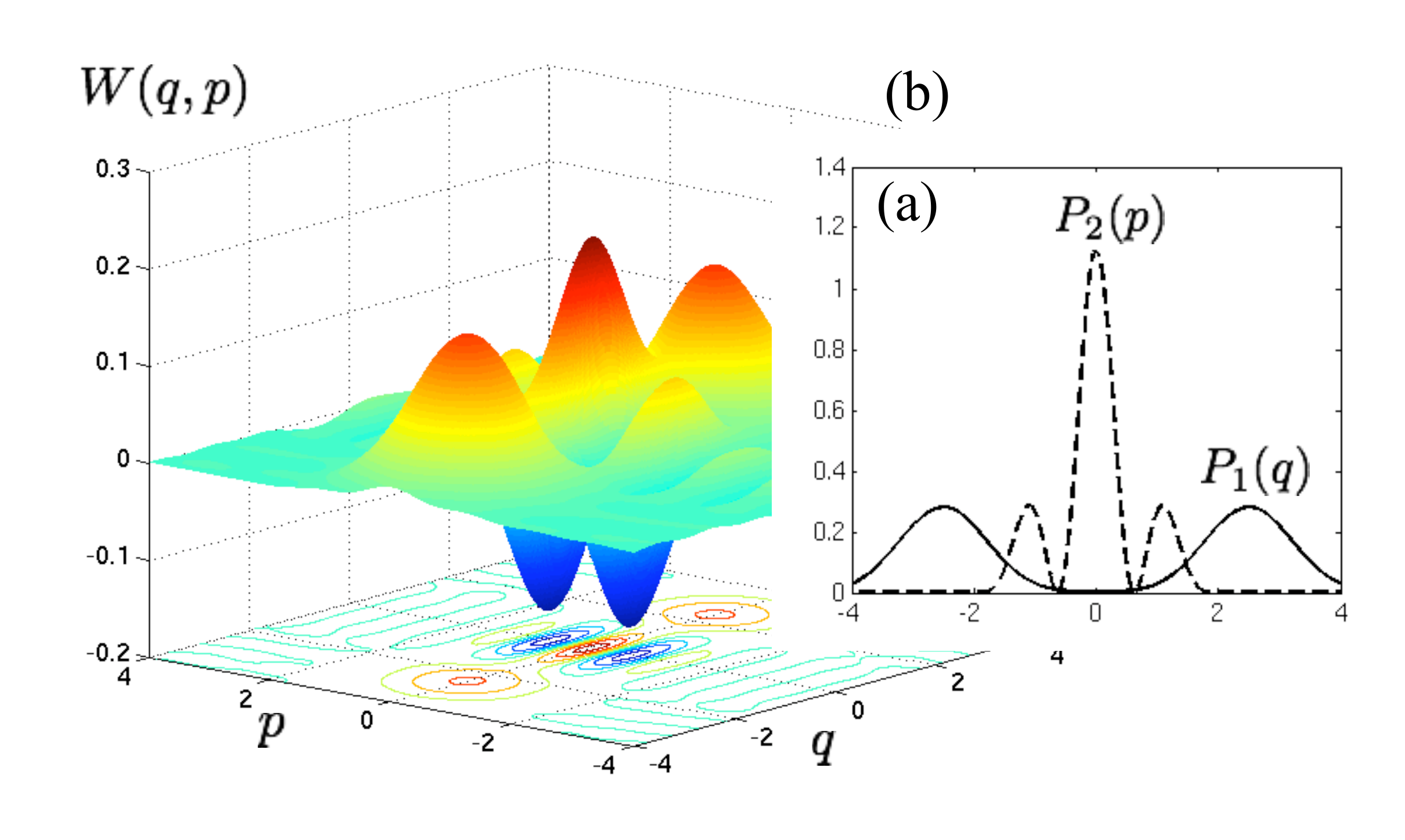}} 
\caption{(a) Position and momentum distributions for an exact cat state. (b) Wigner function
based on the output of the SDP.} \label{cat}
\end{figure}

{\it Characteristic functions and Bochner's theorems. --}
We consider physical systems of $n$ bosonic modes, associated with
canonical coordinates $ R=( q_1,\dots,  q_n,  p_1,\dots,  p_n)$, of ``position''
and ``momentum'', or some
quadratures. 
In the center of the analysis will be quantum characteristic functions
\cite{CVReviews,Leonhardt}. 
For $n$ modes, the {\it quantum characteristic function}
$\chi:\rr^{2n}\rightarrow\cc$
is defined as
\begin{equation}
	\chi(\xi)=\mathrm{tr}[ \rho  D(\xi) ] , \quad    D(\xi)=e^{i\xi \cdot
\sigma  R},
\label{char}
\end{equation}
so as the expectation value of the {\it Weyl} or {\it displacement operator},
where the matrix
\begin{equation}
	\sigma=\left[\begin{array}{cc}
	0& \id_n\\
	-\id_n & 0
	\end{array}\right]
\end{equation}
reflects the canonical commutation relations. 
This characteristic function is nothing but the Fourier transform of the
familiar {\it Wigner function} $W:\rr^{2n}\rightarrow \rr$, given by
\begin{equation}
W(z)=\frac{1}{(2 \pi)^{2n}} \int \chi(\xi)e^{-i\xi \cdot \sigma z} d\xi.
\end{equation}
A key tool in the argument will be the notion of {\it $\lambda$-positivity}
of a phase space function \cite{Hudson}.

\begin{definition}[$\lambda$-positivity]\label{Sigma}
A function $\chi:\rr^{2n}\rightarrow \cc$ 
is said to be $\lambda$-positive definite for $\lambda\in \rr$
if for every $m \in \nn$ and for every set of real vectors
$T= (\xi_1,\xi_2,\dots,\xi_m )$ the  $m\times m$ 
matrix $M^{(\lambda)}(\chi,T)$
with entries
\begin{equation}
 	\left(M^{(\lambda)}(\chi,T)\right)_{k,l}=\chi(\xi_k-\xi_l) e^{i \lambda\xi_k \cdot \sigma \xi_l 
/2}, \label{lambda-p}
\end{equation}
is  non-negative, so $M^{(\lambda)}(\chi,T)\geq 0$. 
\end{definition}

Eq.\ (\ref{char}) defines the characteristic function of a given quantum state.
In turn, one can ask for a classification of all functions that can be characteristic functions
of a quantum state, or some probability distribution in the classical case. 
Such a characterization is captured in the quantum and classical Bochner's
theorems \cite{Hudson}. The following assertions follow: 

(i) Every characteristic function of a quantum state must be {\it 1-positive
definite}.

(ii) Every characteristic function of a quantum state with a positive Wigner
function must be at the same time
{\it 1-positive definite} and {\it 0-positive definite}.

{\it Measuring non-classicality. --} 
Data are naturally taken as slices in phase space, resulting from
measurements of some linear combinations of the
canonical coordinates, as they would be obtained from a phase sensitive
measurement such
as homodyning in quantum optics. 
One collects data from measuring observables
$u_k \cdot R$ for some collection
of $u_k\in \rr^{2n}$ with $\|u_k\|=1$. 
For example, in the simplest case of one mode one could measure only $q$ and $p$
or, if the state is phase invariant, one could average over all the possible
directions.
With repeated measurements one can estimate the associated probability
distributions $P_k:\rr \rightarrow \rr^+$, related to slices of the
characteristic functions by a simple Fourier transform
\begin{equation}
	\int P_k(s)e^{i \omega s}ds
	=\chi(\omega  \sigma u_k). \label{fourier}
\end{equation}
Actually, in a real experiment one can build only a statistical histogram rather
than a continuous probability distribution. As a consequence, measurements of
values of the characteristic function must be equipped with honest error bars. 
An estimate of this error is given by (see Appendix)
\begin{equation}
\delta(\omega)=|\omega| h+{n}/{\sqrt{N}} , \label{error}
\end{equation}
where $2 h$ is the width of each bin of the histogram, $N$ is the number of
measurements and $n$ is the number of standard deviations that one should
consider depending on the desired level of confidence \cite{tube}.
This kind of measurements can be performed also in {\it opto-mechanical systems} where
a particular quadrature of a mechanical oscillator can be measured \textit{a posteriori} by appropriately 
homodyning a light mode coupled to the mechanical resonator \cite{Miao}. 
A different idea has recently been proposed for directly pointwise measuring the characteristic 
function of a mechanical mode coupled to a two-level system \cite{Meystre}. 
In both cases the method that we are going to describe can be easily applied. Restricted measurements
also arise in the context of {\it bright beams} \cite{Leuchs}, 
where Mach-Zehnder interferometers have to replace homodyning in the 
absence of the possibility of having a strong local oscillator. In the study of 
states of {\it macroscopic atomic ensembles} \cite{Mac} similar issues arise.

{\it Bounds to the non-classicality from convex optimization. --} 
We assume that we estimate the values of the characteristic function 
$\chi(\bar \xi_j)\simeq c_j$
for a given set of points $\bar \xi_j$, $j=1,\dots ,p$,
within a given error $\delta_j\ge
0$ \cite{tube}, so that  for all $j$
\begin{equation}
	|\chi(\bar \xi_j)-c_j|\le\delta_j. 
	\label{measurement}
\end{equation} 
Now pick a set of suitable test vectors
$T= (\xi_1,\dots,\xi_m)$, the differences $\xi_j-\xi_k$ 
of which at least contain the 
data points $\bar \xi_1,\dots, \bar \xi_p$.
Based on this, we define the following
convex optimization 
problem as a minimization over $\chi$, $ x$,
\begin{eqnarray}
 \mathrm{minimize} &\; & x,  \label{SDP}\\
 \mathrm{such\,\, that}
 &\; & |\text{Re}(\chi(\bar \xi_j))-\text{Re}(c_j)|\le\delta_j, \, j=1,\dots ,p,
  \label{c1}    \\
 &\; &|\text{Im}(\chi(\bar \xi_j))-\text{Im}(c_j) |\le\delta_j,\, j=1,\dots ,p, 
   \label{c2}    \\
  &\; &M^{(0)}(\chi,T) +  x \,m \id \ge 0,  \label{0-const}    \label{c3}  \\  
  &\; &M^{(1)} (\chi,T)\ge 0,      \label{c4}
\end{eqnarray}
where $M^{(0)}(\chi,T) $ and $M^{(1)} (\chi,T)$ are the Hermitian matrices (\ref{lambda-p})
associated with the $\lambda$-positivity, based on the test points
$\xi_1,\dots,\xi_m$ as being specified in Def.\ \ref{Sigma}.
The minimization is in principle performed over all  functions
$\chi:\rr^{2n}\rightarrow\cc$ 
such that $\chi(-\xi)=\chi(\xi)^*$, where $\chi(\bar \xi_l)$ is constrained by the 
data and $M^{(0/1)}(\chi,T)$ depend on the test points.
Since we take only a
finite number of points of $\chi$, yet,
the above problem gives rise to a {\it semi-definite problem (SDP)} \cite{Convex}. 
This can be efficiently solved with standard numerical algorithms. What is more,
by means of the notion of Lagrange duality, one 
readily gives analytical certifiable bounds to the optimal objective value:
Every solution for the dual problem will
give a proven lower bound to the primal problem  \cite{Convex}, and hence  a
lower bound to the measure of non-classicality
itself. The entire procedure hence amounts to an arbitrarily tight
{\it convex relaxation} of the Bochner constraints.
We can now formulate the main result: Eq.\ (\ref{SDP}) gives rise to a
lower bound for the non-classicality: Given
the data (and errors), one can find good and robust bounds to the 
{\it smallest non-classicality that is consistent with the data}. 

\begin{theorem}[Estimating non-classicality] The output $x'$ of Eq.\
(\ref{SDP}) is a lower bound for the non-classicality, 
$ \eta( \rho)\ge x'$.
\end{theorem}

The proof proceeds by constructing a {\it witness operator}
\begin{equation}
  F=\frac{1}{m} \sum_{k,l=1}^{m} v^*_k v_l D(\xi_k-\xi_l), \label{witness}
\end{equation}
where $\xi_1,\dots, \xi_m\in \rr^{2n}$
are the test vectors from Bochner's theorem used in the SDP
and $v$ is the normalized eigenvector associated with the
minimum eigenvalue of $M^{(0)}(\chi',T)$, where $\chi'$ is the optimal solution for $\chi$. 
For a given state $\rho$, this operator $F$ has the following properties:

1. $F=F^\dag$,

2. $|\text{tr}{( F  \omega)}|\le1$ for all quantum states $ \omega$.

3. $\text{tr}{( F  \omega)}\ge 0$ for all quantum states $ \omega \in
\mathcal C$. 

4. If $x' \ge 0$ is the optimal solution, then 
$\text{tr}{( F \rho)}\le-x'$.

This can be seen as follows: 1.\ follows directly from construction, noting that $D(\xi)=D(-\xi)^\dagger$
for $\xi\in \rr^{2n}$. 
Property 2.\ follows from $|\text{tr}( D(\xi)  \omega)|=|\chi(\xi)| \le 1$,
valid for every characteristic function, from which we find
\begin{eqnarray*}
	|\text{tr}{( F  \omega)}|\le \frac{1}{m} \sum_{k,l=1}^{m} |v_k|  |v_l |=\frac{1}{m} \left( \sum_{l=1}^{m} |v_l |\right)^2\le1,
\end{eqnarray*}
where we have used that for every normalized vector
$ \sum_{l=1}^m |v_l|\leq m^{1/2}$, as can easily be seen by solving a quadratic problem.
To show 3., we can exploit the property of the characteristic function of
$\omega$ of being 0-positive definite. This implies that 
\begin{eqnarray}
	\text{tr}{( F  \omega)}=\frac{1}{m} \sum_{k,l=1}^{m} v^*_k v_l 
	\left(
	 M^{(0)}(\tilde \chi,T) 
	\right)_{k,l}\ge 0,
\end{eqnarray}	
where $M^{(0)}(\tilde \chi,T) $ is the Bochner matrix associated with the characteristic function 
$\tilde\chi$ of  $\omega$.
Finally,
4.\  originates from 
the structure of the SDP as a convex optimization problem.
For optimal solutions $x'$ and $\chi'$, the constraint
(\ref{0-const}) implies that  the minimum
eigenvalue of $M^{(0)}(\chi',T)$ is equal to $-x' m$.
Then $\text{tr}{( F  \rho)}\le \sum_{k,l=1}^{m} v^*_k v_l M_{k,l}^{(0)}(\chi',T)/m=
-x'$ .

The four properties suggest that $F$ is actually a \textit{witness observable}
able to distinguish a subset of 
non-classical states from the convex set of classical states.
Formally, from the variational definition of the trace norm, we have
\begin{equation}
 \eta( \rho)=\min_{ \omega \in \mathcal C}\| \rho- \omega \|_{1}\ge\min_{ \omega
\in \mathcal C} \text{tr}( \omega F) -\text{tr}{( F  \rho)}\ge x',
\end{equation}
which is the lower bound to be shown.

{\it An example: Schr\"odinger cat state. --}
As an example we consider a quantum superposition of two coherent states, so 
$| \psi \rangle\sim (| \alpha \rangle +| -\alpha \rangle )$
with $\alpha=1.77$.
We assume to measure only the probability distributions of
position and momentum (Fig.\ \ref{cat}.a): $P_1(q)= | \langle  q| \psi 
\rangle|^2$ and $P_2(p)=| \langle  p| \psi  \rangle|^2$, that is the data is collected from a mere
{\it pair of canonical operators}.
This amount of information is of course not sufficient for tomographically 
reconstructing the
state since it corresponds to just two orthogonal slices of the characteristic
function.

In order to define the SDP we consider a $25 \times 25$ square lattice centered
at the origin of the domain of the characteristic function and we optimize over
the values of $\chi$ at the lattice points. Position and momentum measurements
can be used to define constraints (\ref{c1}-\ref{c2}) for only two slices of the
lattice and we assume an error of $\delta_j=10^{-3}$ for each point. We generate
$100$ random test vectors and we construct the associated $\lambda-$positivity
constraints (\ref{c3}-\ref{c4}).
The output of the SDP is $x' \simeq 0.05>0$ which is a certified lower
bound for the non-classicality of the state.

\begin{figure}[tbh]
\centerline{\includegraphics[width=0.42\textwidth]{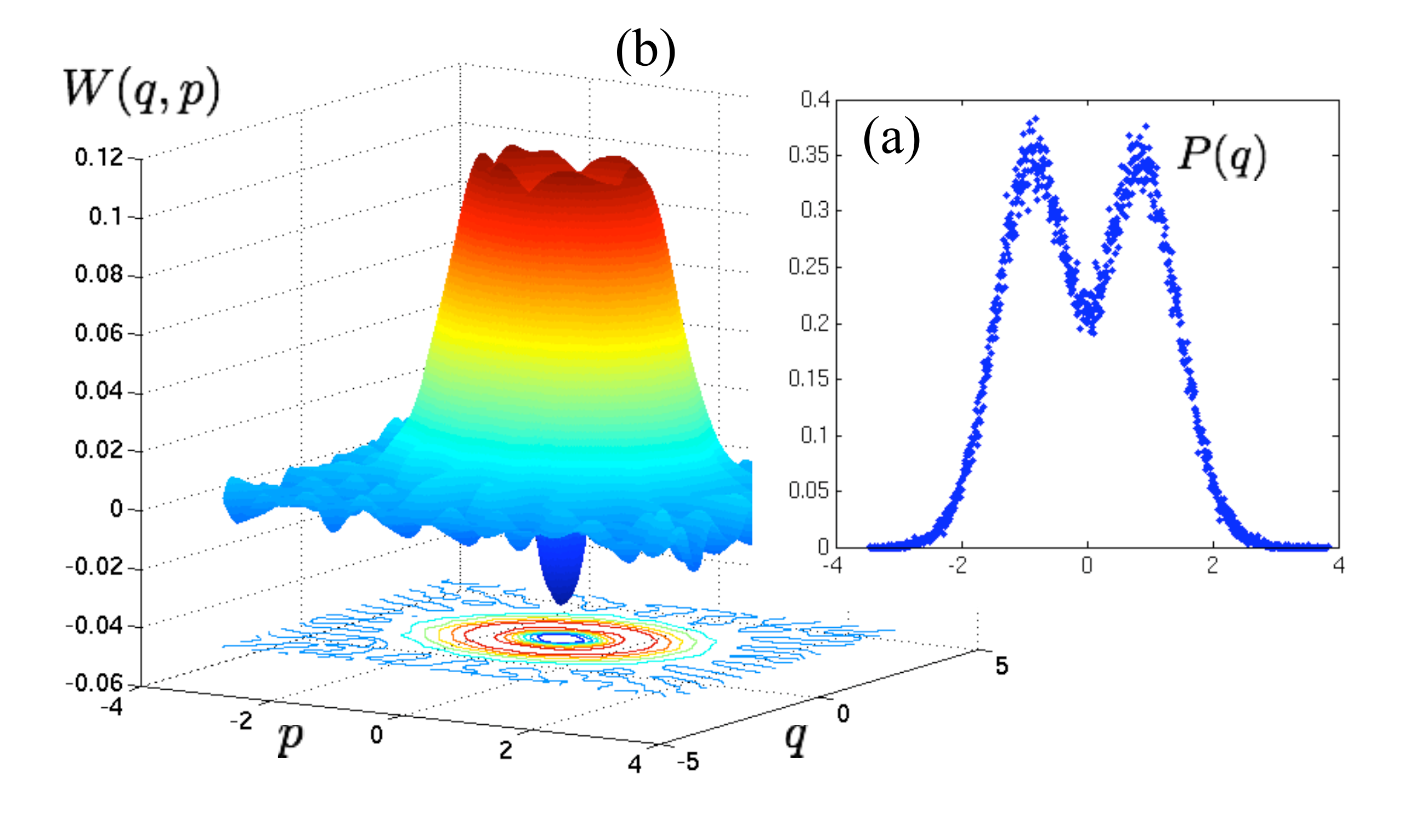}} 
\caption{(a) Raw measured quadrature distribution from the experiment
without any correction. (b) Related Wigner function 
based on the output of the SDP.} \label{fock}
\end{figure}

{\it Experimentally detecting non-classicality. --} 
Finally, to certify the functioning of the idea in a quantum optical context,
we apply our method to experimental data.
We consider data from a heralded {\it single-photon source based on parametric 
down-conversion} (cf.\ Ref.\ \cite{Polzik2}). 
Here, one of the two down-converted photons is detected and heralds the presence of the other photon. The heralded single photon is sampled $180,000$ times by homodyne measurements at phase-randomized quadratures (Fig.\ \ref{fock}.a).
Instead of using these data in tomography we will use the same data to directly extract a certified lower bound to the non-classicality of the state.  We will show that the most classical state consistent with those data has a negative Wigner function (Fig.\ \ref{fock}.a) which is close to the one reconstructed with a maximum likelihood method in Ref.\ \cite{Polzik2}. This strengthens the finding of a negative Wigner function, as we hence directly certify it as a worst case bound including error bars.
Data correspond to phase randomized measurements meaning that we can use the
same probability distribution for every phase space direction, the phase being not
available in this experiment. Since our
non-classicality measure is convex, averaging over phase space directions is an
operation which can only decrease the negativity of the state. This means that
a lower bound to the non-classicality of the phase randomized state will  
be valid for the original state.

In order to apply our algorithm we use the measured data to constrain all the
points of the characteristic function on a $37 \times 37$ lattice. Error bars
are estimated using Eq.\ (\ref{error}) with $n=5$ standard deviations. This means
that the probability that all the points of the lattice lie inside the error
bars is larger than $99.9 \%$. The lower bound for the non-classicality coming
out from the SDP ($200$ random  test vectors have been used) is larger than zero
$x' \simeq 0.0028$, meaning that the Wigner function of the state cannot
be a positive probability distribution.  Indeed, the Wigner
function reconstructed from the optimal solution of the SDP (Fig.\ \ref{fock}.b)
is clearly negative even if we asked for
the most positive one consistent with measured data. 

{\it Extensions of this approach. --} Needless to say, this approach can be 
extended in several ways. Indeed, the method can readily be generalized to produce lower 
bounds for {\it entanglement measures} \cite{Entanglement} in 
the multi-mode setting. Also, this idea can be applied to the situation when not slices 
are measured, but {\it points in phase space}, 
such as when using a detector-atom that is simultaneously coupled to a 
cantilever \cite{Meystre}. It would also constitute an interesting perspective
to see how the present ideas can be used to certify deviations from {\it stabilizer
states} for spin systems, being those states with a positive discrete Wigner function \cite{Gross}.

{\it Summary. --} We have introduced a method to directly measure 
the non-classicality of quantum mechanical modes, requiring
less information than tomographic knowledge, but at the same time in a 
certified fashion. In this way, these ideas are further advocating the 
paradigm of ``learning much from little''---getting much certified information from few measurements---complementing 
methods of {\it witnessing entanglement} \cite{Entanglement,Witness}, recent ideas
of {\it compressed sensing} \cite{Compressed} or {\it matrix-product based} \cite{MPS} approaches 
to quantum state tomography, {\it detector tomography} \cite{Detector}, 
or the direct estimation of {\it Markovianity} \cite{Markovian}
of a continuous process. It is the
hope that this paper further stimulates work in the context of this paradigm.

{\it Acknowledgments. --} This work has been supported by the EU (MINOS,
COMPAS, QESSENCE), and the EURYI.

\section{Supplementary information (EPAPS)}

In this appendix, we add a discussion on measurement errors not essential for the main text.
We assume data to be organized in histograms made of bins of width $2 h$
centered at points $s_j=2 h j$. The normalized height of the $j$-{th} bin is
\begin{equation}
P_k^j =\int_{s_j-h}^{s_j+h} P_k(s)ds \label{histo}+w_j,
\end{equation}
where $\{w_i\}$ are random variables with zero average associated to the
statistical error. In a standard homodyne detection, $\{w_j\}$ have Poissonian
variances equal to $\langle w_jw_l \rangle=\delta_{j,l} 
\int_{s_j-h}^{s_j+h} P_k(s)ds/N$, where $N$ is the total number of measurements.
The simplest approximation for the value of the characteristic function defined in 
Eq.\ (\ref{fourier}) is
\begin{equation}
\chi^a(\omega \sigma u_k)=\sum_{j} P_k^j e^{i \omega s_j}.
\end{equation}
The error will be the sum of two terms
\begin{equation}
\chi(\omega  \sigma u_k ) -\chi^a(\omega  \sigma u_k) = \Delta + \tilde w.
\end{equation}
The first term is the error due to the finite resolution of the histogram, which is bounded
by $|\Delta| \le h |\omega|$, while $\tilde w=\sum_{j} w_j e^{i \omega s_j}$ is the random statistical error.
This error could be estimated and possibly reduced by using \textit{bootstrapping} or \textit{maximum likelihood} algorithms.
For our purposes, we simply calculate the standard deviation of the random variable $\tilde w$, which is given by 
$({\langle |\tilde w|^2\rangle})^{1/2}=1/\sqrt{N}$. Now, for the central limit theorem $\tilde w$ can be approximated by a Gaussian distribution, meaning that the probability that the following tube constraint is violated is exponentially suppressed in the number of considered standard deviations $n$ \cite{tube},
\begin{equation}
|\chi(\omega  \sigma u_k ) -\chi^a(\omega  \sigma u_k)| \le h |\omega|+{n}/{\sqrt{N}}.
\end{equation}


\begin{thebibliography}{99}
\bibitem{tube}
	Since the error follows a Gaussian distribution, the probability $P_{\text{out}}$ that 
	the actual value of the characteristic function lies outside the error bar is never 
	zero but it is exponentially suppressed for increasing $n$. E.g., for $n=3$ 
	one has $P_{\text{out}} < 2.7 \times 10^{-3}$, for $n=5$ one has $P_{\text{out}}<6 
	\times 10^{-7}$, \textit{etc.}.
\end{thebibliography}
\end{document}